\documentclass[multphys,vecphys]{svmult}

\usepackage{makeidx}         
\usepackage{graphicx}        
\usepackage{multicol}        
\usepackage[bottom]{footmisc}

\makeindex             

\begin{document}

\title*{Cold fronts in cool core clusters}
\author{Simona Ghizzardi \and
Silvano Molendi \and Mariachiara Rossetti \and Alberto Leccardi}
\authorrunning{Ghizzardi et al. } 
\institute{IASF Milano, INAF
\texttt{simona@iasf-milano.inaf.it}}
%
%
\maketitle

\begin{abstract}
Cold fronts have been detected both in merging 
and in cool core clusters, where little or no sign of a merging event is present.
A systematic search of sharp surface brightness
discontinuities performed 
on a sample of 62 galaxy clusters observed with XMM-Newton shows that
cold fronts are a common feature in galaxy clusters.  Indeed most
(if not all) of the nearby clusters ($z < 0.04$) host a cold front.
Understanding the origin and the nature of a such frequent phenomenon is
clearly important. 
To gain insight  on the nature of cold fronts in cool core
clusters we have undertaken a
systematic study of all contact discontinuities detected in our sample,
measuring surface brightness, temperature and when possible
abundance profiles across the fronts. 
We measure the Mach numbers for the cold fronts
finding values which range from 0.2 to 0.9; we also
detect a discontinuities in the metal 
profile of some clusters.  
\end{abstract}

\section{Introduction}
\label{sec:intro}

Chandra and XMM-Newton observations have shown that 
many clusters host very sharp surface brightness discontinuities.
The drop in the X-ray surface brightness
is accompanied by a rise of similar magnitude in the gas temperature.
These discontinuities, dubbed {\it cold fronts} \cite{Vik:2001},
have been detected both in merging and in cool core clusters. 
While the first ones have been immediately related to the merging 
event \cite{Maxim:2000}, the origin of the latter is still not 
fully understood. 
As we will show, cold fronts are quite common in cool core clusters. Thus, 
understanding the nature of such a widespread phenomenon is 
mandatory to characterize the dynamics of galaxy clusters and of their cores. 

The most popular scenarios to solve the cooling 
flow problem concern AGN and the interaction between the 
radio lobes inflated by the AGN and the ambient gas itself.
The mechanical energy transferred from
the lobes to the thermal plasma is in many, but not all cases,
sufficient to quench the cooling (see \cite{Rafferty:2006}).
 
Another class of heating sources includes mechanisms which 
provide heat from outside the cool core region. In this
context considerable effort has gone into exploring the role of
conduction (i.e. \cite{Voigt:2004}, \cite{Ghizzardi:2004}). 
As it turns out there are various difficulties: firstly
the conductivity of the ICM is unknown; secondly in some systems the
temperature profiles are rather flat (i.e. M87, Ghizzardi et al.
2004 \cite{Ghizzardi:2004}), and thirdly, while heating from conduction scales like
T$^{5/2}$, cooling is more efficient at lower temperatures. 

Another possible way of heating the flow from outside has to do 
with the ubiquitous presence of cold fronts in cluster cores.
Although the kinetic 
energy in these cold fronts, a fraction of the thermal energy, 
is in itself insufficient to offset the cooling, cold fronts could 
act as an energy reservoir. Hydrodynamical simulations 
(\cite{TH2005} and \cite{Asca:2006}) show that cold fronts in cores
could be set off by minor, frequent, mergers that would provide a
constant supply of energy. Clearly a detailed observational
characterization of cold fronts
is of primary importance to explore this scenario in greater detail. 
We study cold fronts in a large sample of galaxy clusters 
observed with XMM-Newton to provide a precise description of this 
phenomenon and 
to better understand the dynamics of cluster cores.

\section{Hunting for cold fronts in a large XMM-Newton cluster sample.}
\label{sec:statistic}

We have performed a systematic characterization of cold fronts in
a sample of 62 clusters observed with XMM-Newton. 
The large collecting area of the 
EPIC telescope onboard the XMM-Newton satellite 
allows a detailed inspection of the spectral properties of the 
galaxy clusters, which are important to study the dynamics of the core.
The sample includes two different subsamples.
The first comprises roughly 20 nearby bright clusters 
with redshifts in the range $[0.01-0.1]$. The second subsample 
comprises all the clusters available in the XMM-Newton public archive
up to March 2005,  
having redshifts in the range $[0.1-0.3]$ (see \cite{Leccardi:2006}).

The systematic search for surface brightness and temperature discontinuities 
in the clusters of our sample resulted in  the detection of cold fronts in 21 objects
corresponding to a percentage of 34\%.
It is interesting to study the frequency of cold fronts in different redshift ranges.
If we progressively reduce the sample, excluding 
gradually the more distant clusters, the fraction of 
clusters having a cold front increases. The occurrence of cold fronts is 41.8\%
for clusters with redshift $0.01<z<0.2$ , 50\% for clusters with $0.01<z<0.1$ 
and 72.2\% for clusters with $0.01<z<0.07$.
A large fraction (87.5\%) of  the nearest clusters ($z < 0.04$) host one 
or more cold front.
%
%
This is in agreement with results derived analyzing a sample 
of 37 relaxed nearby clusters observed with {\sl Chandra} \cite{Maxim:aph}.
Since projection effects and the XMM-Newton resolution 
can hide a non-negligible fraction of 
cold fronts, our result implies that probably all the nearby clusters host 
one (or more) cold fronts.

\section{Characterizing cold fronts}
\label{sec:CF_char}

The results reported in the previous section show that cold fronts are 
indeed a common feature in cluster cores. 
The precise characterization of 
this phenomenon is a preliminary and necessary step in assessing wether  
they can play a significant role in the cooling-heating 
balance within the cluster cores. 
First, we consider some general features of cold fronts in galaxy clusters
(Sec.~\ref{sub:general}); we then characterize the discontinuities by measuring the 
surface brightness  and temperature jump 
(Sec.~\ref{sub:jump}) and finally we 
derive abundance profiles across cold fronts (Sec.~\ref{sub:metal}).

\subsection{General aspects of cold fronts in cool core clusters}
\label{sub:general}

We have derived surface 
brightness, temperature and pseudo-pressure (hereafter SB, T and P) maps 
for some cool core clusters using the 
{\it adaptive binning + broad band fitting} 
method (see \cite{Rossetti06} for a detailed description of this procedure). 
Two relevant features have been detected: a surface brightness 
peak displacement and a spiraling pattern in the temperature maps.
As an example we consider 2A0335+096 \index{2A0335+096}, 
which hosts a cold front in the southern 
direction $\sim 70 ''$ from the SB peak.  
As already outlined in \cite{Ghizzardi:2004}, 
the position of the BCG of this cluster matches the P
peak, while the SB and the T peaks are shifted in the south 
direction by ~16 arcsec.
Similar shifts have been observed in others 
(i.e. A1795) but not in all (i.e. A496) 
clusters.

\begin{figure}
\centering
\includegraphics[height=4.3cm]{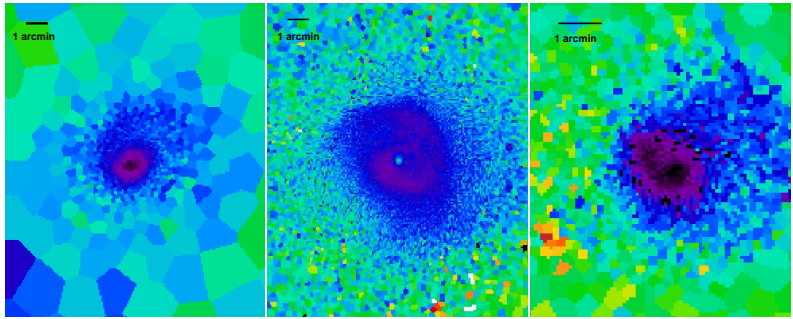} 
%
\caption{Temperature maps for (a) 2A0335+096, (b) Perseus and (c) Centaurus. All of them 
show a spiral pattern.}
\label{fig:spiral}       
\end{figure}

Another interesting feature is visible in the T maps. 
In Fig.~\ref{fig:spiral} we show the T maps for 
2A0335+096 \index{2A0335+096}, Perseus \index{Perseus} and Centaurus \index{Centaurus}:
all of them show a spiral structure. 
Both phenomena have a natural explanation within 
the scenario proposed by \cite{Maxim:aph} and \cite{Asca:2006}. 
According to their model, 
cold fronts are formed when the central cold gas 
is subsonically sloshing in the dark matter 
gravitational potential.  
Frequent minor mergers can induce gas oscillations in the cluster core displacing 
the thermal gas from the bottom of the potential well. 
Under these circumstances, 
the P peak and the BCG which trace the gravitational potential, and the SB and T 
peaks which trace the thermal gas, decouple. 
Simulations by Ascasibar and Markevitch \cite{Asca:2006} also show that 
the gas can acquire some angular 
momentum while oscillating. In this case, 
some spiraling structure in the T map is induced.

\subsection{Measuring discontinuities and velocities of cold fronts}
\label{sub:jump}

\begin{figure}
\centering
\includegraphics[height=9.cm]{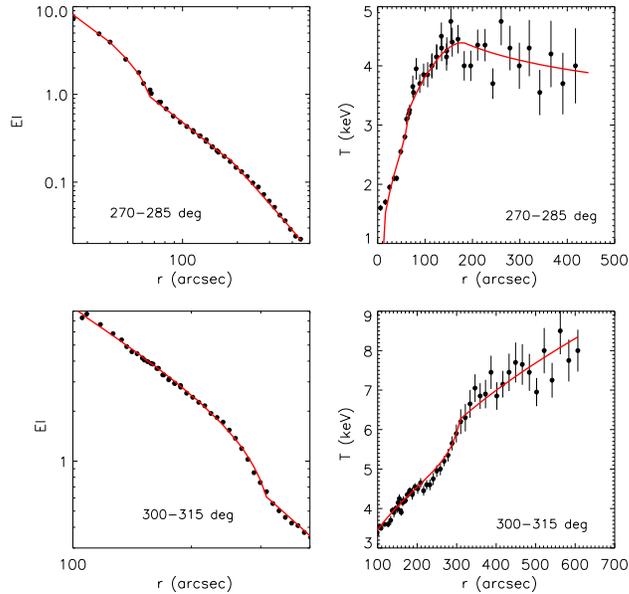}  
%
%
\caption{Surface brightness and temperature profiles for 
2A0335+096 (upper panels) and Perseus (lower panels) with the best fits.}
\label{fig:jump}       
\end{figure}

To quantify the dynamics of cold fronts in cool cores we derive the SB and T 
profiles for each cluster in 15 degrees wide sectors. The sectors have been chosen 
following the SB contour levels in order to properly characterize the jump.
Classic deprojection procedures cannot be applied because of the lack of spherical 
symmetry.
We describe the cold front discontinuity by modeling the electronic density and the gas 
temperature with power laws inside and outside the cold front edge. We then  
project these quantities  assuming that the cold front has 
a width of $\Delta\varphi$ and an inclination angle $\Delta\vartheta$ 
with respect to the line of sight. 
A detailed description of this procedure will be given in \cite{Ghizzardi:06}.
In Fig.~\ref{fig:jump} we plot as dots the SB and the projected T for  the southern 
sector in 2A0335+096 
(top panels) and for the southwestern sector in Perseus
(bottom panels);
the solid lines  are the best fits. For 2A0335+096, 
we find a density jump $n_{in}/n_{out}=1.97$  and a temperature jump 
$T_{in}/T_{out}= 0.88$.  This 
corresponds to a Mach number of $\cal{M} $ $= 0.86$. 
For Perseus we find a density jump $n_{in}/n_{out}=2.1$  and a temperature jump    
$T_{in}/T_{out}= 0.79$ corresponding to a 
Mach number $\cal{M}$ $= 0.8$ . 
A more complete and detailed analysis of the discontinuities for all the clusters of our 
sample will be presented in \cite{Ghizzardi:06}.

\subsection{Metal profiles across the cold fronts}
\label{sub:metal}

\begin{figure}
\centering
\includegraphics[height=4.cm]{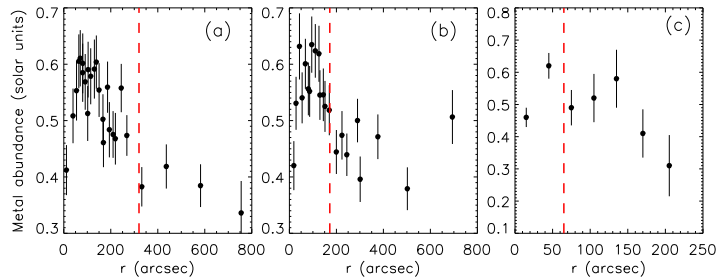} 

%
\caption{Metal profiles in some sectors of Perseus (a-b) and 2A0335+096 (c). 
The dashed lines 
mark the cold fronts positions. A discontinuity is clearly visible in the Perseus cluster. 
Some indication of discontinuity can be seen for 2A0335+096. }
\label{fig:metal}       
\end{figure}

We derive the  iron abundance profiles across the cold front for some clusters. 
Fig.~\ref{fig:metal} shows the profiles for Perseus (northern and southwestern sectors) 
and for 2A0335+096 (southern sector). The dashed lines mark the position 
of the cold fronts. 
The iron abundance has a discontinuity across the cold fronts for the Perseus cluster. 
Even if the data quality for 2A0335+096 is not as high as for Perseus, an indication for a 
discontinuity seems to be present also for this cluster. This behavior is expected 
in a sloshing scenario as the cold metal rich gas is shifted towards more 
external regions where
the iron abundance is lower.

%
%
\section{Conclusions}
\label{sec:concl}

Cold fronts could play an important role in providing heating  
from the outer regions of the core 
giving a contribution to quenching the cooling flow in cool cores.
The main advantage in considering such phenomenon as a possible heat source is twofold: 
cold fronts are a common feature in the galaxy cluster population;
they could have a purely "gravitational origin".
Ascasibar and Markevitch \cite{Asca:2006}
show that frequent minor mergers can induce a disturbance to the gravitational 
potential well and cause gas sloshing and cold fronts. 
Each cluster during its formation undergoes several 
minor merger events. Among the possible heating processes, 
this mechanism has the advantage of being
common to all the clusters and of being
unrelated to any particular feature of the cluster 
(i.e. the presence of an active AGN.)

An analysis of a large sample of clusters observed with XMM-Newton shows that 87.5\% 
of the nearby ($z < 0.04$) clusters host a cold front.
The analysis of the surface brightness, 
temperature and pressure maps for some clusters 
shows that in some cases the SB and T peaks are decoupled  with respect 
to the P peak. In some clusters, we also observe a spiraling 
pattern in the temperature map; both these features are 
expected in a sloshing scenario \cite{Asca:2006}.

We have measured the Mach numbers of cold fronts, 
finding values that range from 0.2 to 0.9.
The analysis of the metal profiles shows that the iron abundance has a 
sharp discontinuity  across the cold front edge in Perseus 
and some indications of that are present also for 2A0335+096. 
A detailed characterization of the discontinuities can help to understand the 
dynamics of the innermost regions of the cluster and to 
quantify the amount of heating that the cold front can provide to cluster cores.

%
%
%

%
%



\printindex
\end{document}